\documentclass[11pt]{article}

\usepackage{amssymb,amsfonts,amsmath,amsthm}
\usepackage{apacite} 
\usepackage{bm}
\usepackage{booktabs}
\usepackage{cases}
\usepackage{csquotes}
\usepackage{cleveref}
\usepackage{comment}
\usepackage{color, colortbl}
\usepackage[font={small},width=.9\textwidth]{caption}
\usepackage{enumerate}
\usepackage{fancyhdr} 
\usepackage[margin=1in]{geometry}
\usepackage{graphicx,rotating}
\usepackage{indentfirst}
\usepackage{longtable}
\usepackage{mathrsfs}
\usepackage{mwe,tikz}
\usepackage{mathtools}
\usepackage[elide]{natbib}
\usepackage[percent]{overpic}
\usepackage{pslatex} 
\usepackage{pdflscape}
\usepackage{picture,xcolor}
\usepackage{setspace}
\usepackage{subcaption}
\usepackage{titlesec}
\usepackage{tocloft}
\usepackage{units}

\usepackage{apalike} 

\theoremstyle{plain}
\numberwithin {equation}{section}

\newtheorem{lemma}{Lemma}[section]

\theoremstyle{example}

\theoremstyle{definition}

\theoremstyle{remark}

\newcolumntype{R}[1]{>{\raggedleft\let\newline\\\arraybackslash\hspace{0pt}}m{#1}}

\newcommand{\Sn}{S_{n}}

\newcommand{\cov}{\mathrm{Cov}}

\newcommand{\lam}{\lambda}

\newcommand{\Sig}{\bm{\Sigma}}
\newcommand{\X}{\mathbf{X}}

\newcommand{\U}{\mathbf{U}}

\newcommand{\Lam}{\bm{\Lambda}}

\newcommand{\hatlam}{\hat{\lambda}}

\newcommand{\Sbf}{\mathbf{S}}

\newcommand{\Z}{\mathbf{Z}}

\newcommand{\MP}{F_{y,H}}

\newcommand{\MPm}{m_{y,H}}

\newcommand{\MPpsi}{\psi_{y,H}}
\newcommand{\MPsupp}{\Gamma_{F_{y,H}}}
\newcommand{\Hsupp}{\Gamma_{H}}

\newcommand{\PSD}{H(t; \bm{\theta})}

\graphicspath{{figures/}}

\begin{document}
\title{Estimation of the number of spikes using a generalized spike population model and application to RNA-seq data}
\author{Hyo Young Choi and J. S. Marron\\
Department of Statistics and Operations Research, UNC Chapel Hill}
\date{}
\maketitle


\textbf{Abstract.} Although a generalized spike population model has been actively studied in random matrix theory, its application to real data has been rarely explored. We find that most methods for determining the number of spikes based on the Johnstone's spike population model choose far too many spikes in RNA-seq gene expression data or often fail to determine the number of spikes by indicating that all components are spikes. In this paper, we propose a new algorithm for the estimation of the number of spikes based on a generalized spike population model. Also, we suggest a new noise model for RNA-seq data based on population spectral distribution ideas, which provides a biologically reasonable number of spikes using the proposed algorithm. Furthermore, we propose a graphical tool for assessing the performance of the underlying noise model. 
~\\


\thispagestyle{empty}
\bigskip
\pagebreak
\setcounter{page}{1}
\section{Introduction}

Principal component analysis (PCA) is an important dimension reduction tool which finds a low dimensional subspace maximizing the explained variation in data. In high dimensional settings, a collection of observations can be thought of as a linear combination of a small number of source signals plus noise, and PCA can be used to approximately recover the unknown signals. That is, each observation vector $X_j$ can be modeled by
\begin{eqnarray}\label{eq:signal_model}
	X_j = \mu + Ay_j + \epsilon_j 
\end{eqnarray}
where $\mu$ is a mean vector, $A$ is a $d \times K$ matrix representing source signals in its columns, $y_j$ is an $K$-dimensional random vector, and $\epsilon_j$ is a $d$-dimensional vector of noise. Recent work based on this model (\ref{eq:signal_model}) includes \cite{kritchman2008determining,passemier2012determining,ma2013sparse,shabalin2013reconstruction,choi2014selecting,yao2015large,fan2015asymptotics}. For instance, in a signal detection model, $X_j$ can be a vector of the recorded signals with noise at a certain time, where the columns of $A$ are $K$ unknown source signals, and the $y_j$'s are emission levels of these signals. See e.g. Section 11.6 of \cite{yao2015large}. In econometrics, $X_j$ can be the returns of stocks at a certain time, where A is a matrix of latent common factors and the $y_j$'s are unobservable random factors \citep*{onatski2012asymptotics,ma2013sparse,fan2015asymptotics}. From now on, without loss of generality, we assume that the mean vector is zero, i.e. $\mu=0$, by subtracting the sample mean.\\
\indent In many related works, the $d$-dimensional noise vector $\epsilon_j$ in (\ref{eq:signal_model}) is modeled by $\epsilon_j = \sigma z_j$ where $\sigma>0$ is the noise level and $z_j$ is a $d$-dimensional vector of i.i.d. white noise. Also, it is often assumed that $y_j$ and $z_j$ are independent. Then, the covariance matrix of $X_j$ becomes
\begin{eqnarray*}
	\Sig = A\cov(y_j)A^{T} + \sigma^2 I_d.
\end{eqnarray*}
Let $\tilde{\alpha}_1, \cdots, \tilde{\alpha}_K$ denote the eigenvalues of $A\cov(y_j)A^{T}$. Since the rank of $A\cov(y_j)A^{T}$ is at most $K$, the eigenvalues of $\Sig$ are 
\begin{eqnarray}\label{eq:spectrum}
	\U^{T}\Sig \U = \Lam =  (\underbrace{ \tilde{\alpha}_1 + \sigma^2, \tilde{\alpha}_2 + \sigma^2, \cdots, \tilde{\alpha}_K + \sigma^2}_{K} , \underbrace{\sigma^2 , \cdots, \sigma^2}_{d-K})
\end{eqnarray}
where $\tilde{\alpha}_1 \geq \tilde{\alpha}_2 \geq \cdots \geq \tilde{\alpha}_K \geq 0$ and the $\Sig$ has the spectral decomposition $\Sig = \U \Lam \U^{T}$. In this paper, we denote the large eigenvalues in (\ref{eq:spectrum}) by $\alpha_i = \tilde{\alpha}_i + \sigma^2,~i=1, \cdots, K$ for convenience, i.e.
\begin{eqnarray}\label{eq:spike-model}
	\U^{T}\Sig \U = \Lam = \mathrm{diag}(\alpha_1, \cdots, \alpha_K, \sigma^2, \cdots, \sigma^2)
\end{eqnarray}
with $\alpha_1 \geq \cdots \geq \alpha_K > \sigma^2 > 0 $. This model is called the \textit{spiked covariance model} \citep*{johnstone2001distribution}. The eigenvalues $\alpha_1, \cdots, \alpha_K$ correspond to spikes and they are mostly assumed to be much larger than the non-spike eigenvalues. The directions corresponding to spikes can be considered as important underlying structure of the data set whereas the other directions corresponding to non-spikes can be considered as noise. 

One challenging problem in PCA is how to determine the number of spikes $K$. The scree plot is one of the popular ways that have been proposed. Based on the plot of ordered eigenvalues, one looks for an elbow that distinguishes the eigenvalues that are remarkably large from relatively small eigenvalues. Although this method is very simple, looking for an elbow is often subjective and it may be unclear that the elbow really gives the meaningful separation. Moreover, when one has a large number of datasets to be analyzed, it is hard to look at all scree plots corresponding to each dataset. For example, when applying PCA to RNA-seq data for many separate genes which is a main interest in the paper, visual inspection would entail looking at more than 20,000 scree plots and deciding on elbows. This is obviously intractable.

\begin{figure}[t]
	\centering
	\begin{overpic}[width=15cm,page=6]{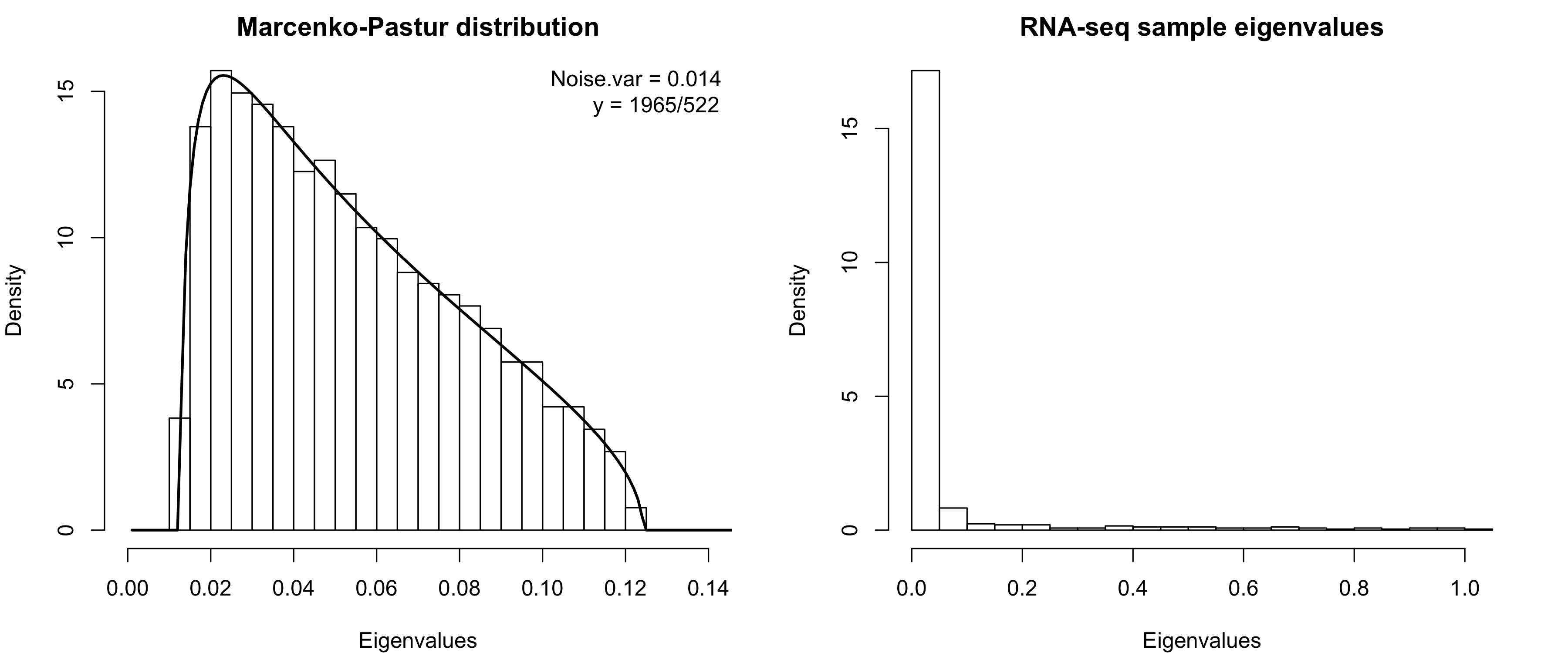}
	\put(75,19){\colorbox{white}{\fbox{\includegraphics[width=3cm,page=1,trim=3.6in 2in 2in 1.5in,clip]{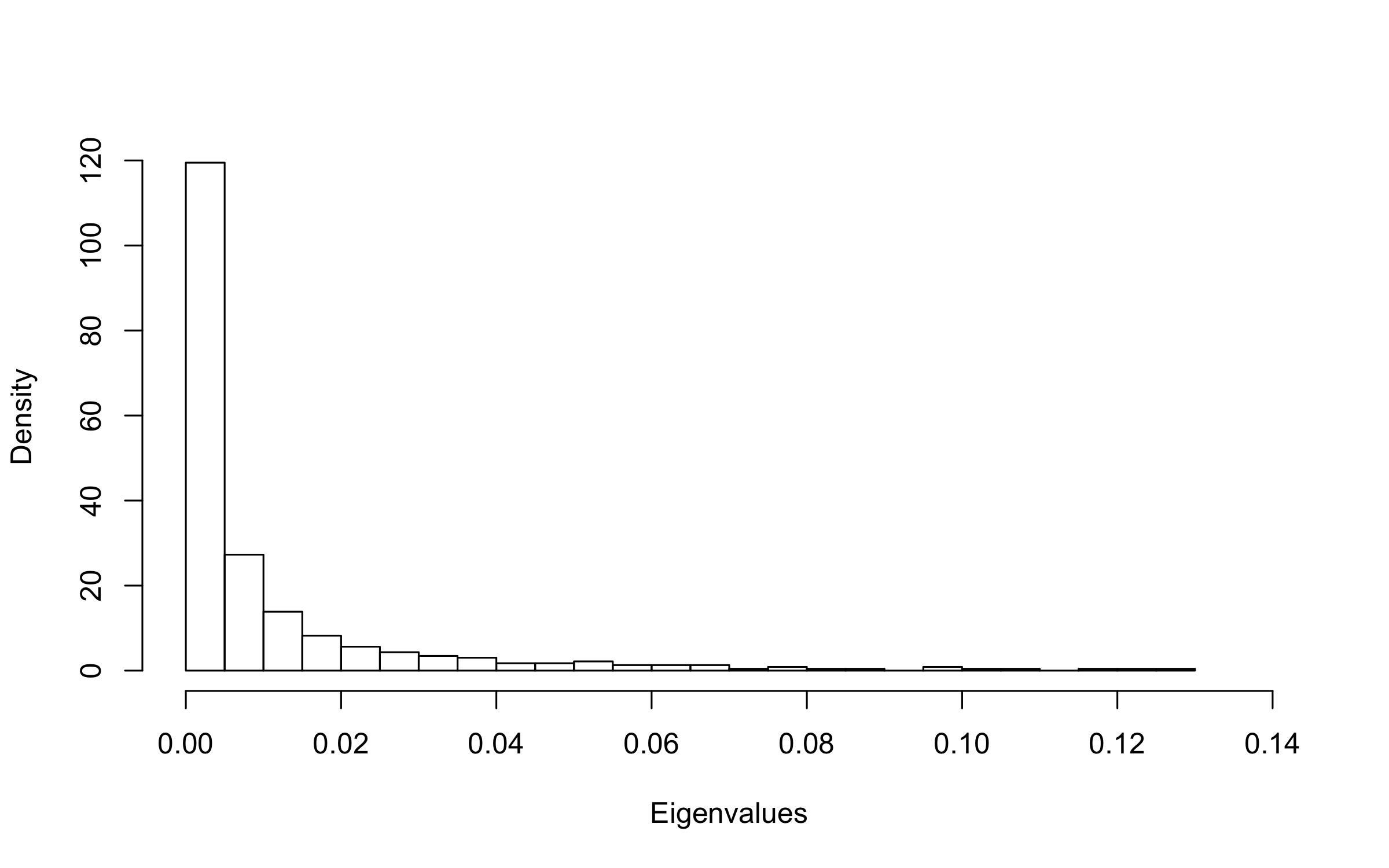}}}}
    \end{overpic}
	\caption{The left plot shows the histogram of simulated eigenvalues under a white noise assumption with $d=1985$ and $n=522$ where the $d$ and $n$ are consistent with the RNA-seq data at the gene CDKN2A. The curve shows the theoretical Marcenko-Pastur distribution with parameters $\sigma^2 =0.014$ and $y=d/n$. The $\sigma^2$ is estimated from the sample eigenvalues. The right plot shows the histogram of the RNA-seq sample eigenvalues excluding a few large ones. Inset box shows the same zoomed-in histogram using the range shown on the right panel. The figure shows that these distributions are widely different.}
\label{fig:MPdist}
\end{figure}

Methods for determining the number of spikes in PCA have been developed in different fields including statistics \citep*{besse1993application,krzanowski1995cross,choi2014selecting}, signal processing \citep*{wax1985detection,kritchman2008determining, kritchman2009non}, and econometrics \citep*{harding2007structural,passemier2012determining,passemier2014estimation}. For excellent background, see Chapter 6 of \cite{jolliffe2002}. As a related problem of a low rank matrix construction based on SVD has been studied in \cite{wongsawat2005multichannel,shabalin2013reconstruction,nadakuditi2014optshrink}. Most of the previous works assumed that the population eigenvalues corresponding to non-spikes are all equal as in (\ref{eq:spike-model}). Under (\ref{eq:spike-model}), the empirical distribution of non-spike sample eigenvalues, that is the sample eigenvalues except for a few first large ones, should be close to the classical Marcenko-Pastur (M-P) distribution, which is the limiting spectral distribution (LSD) of $\Sbf_n = \frac{1}{n}\sum_{j=1}^{n}X_j X_j^{T}$ when $d$ and $n$ both tend to infinity such that $\frac{d}{n}\rightarrow y$. However, as in Figure \ref{fig:MPdist}, we have observed that non-spike sample eigenvalues from RNA-seq data do not follow the classical M-P distribution. They rather show even more heavy tails and right skewness. Figure \ref{fig:MPdist} compares the empirical spectral distribution (ESD) from the RNA-seq data at the gene CDKN2A after excluding a few extremely large eigenvalues with the theoretical M-P distribution based on $y=\frac{d}{n}$ where $d$ and $n$ are the same dimension and sample sizes of the data. The noise variance, $\sigma^2$ is estimated using the remaining sample eigenvalues. For better comparison, the ESD on the same x-axis as the M-P distribution is also given in a small box. Clearly, the ESD from the data is extremely right-skewed and this is even true when we remove more large eigenvalues. As shown in Section \ref{section:application}, due to such extreme positive skewness in the spectral distribution, most existing methods choose hundreds of spikes in RNA-seq data or often fail to determine the number of spikes by providing all PCs as spikes. In Section \ref{section:application}, we show that the white noise model that is assumed in the existing methods is not appropriate for the RNA-seq data and propose a new noise model which successfully models the ESD of the data providing even better fit by capturing the extreme right-skewness. Furthermore, we propose a new algorithm based on the proposed noise model, which provides a reasonable number of principal components (PC) and thus enables us to use the chosen PCs in downstream statistical analyses.  

The remainder is organized as follows. Some relevant known results in random matrix theory are reviewed in Section \ref{section:RMT}. Section \ref{section:method} describes our methods based on the theoretical results. In Section \ref{section:application}, we apply the proposed method to RNA-seq data and compare the results with some existing methods.

\section{Known results on the generalized spike covariance model}\label{section:RMT}

In this section, we review some important results of random matrix theory. Let $\{w_{ij}\}_{1\leq i \leq d, 1 \leq j \leq n}$ be i.i.d. random variables satisfying
\begin{eqnarray*}
E(w_{11})=0, ~~E(|w_{11}|^2)=1, ~~E(|w_{11}|^4) < \infty
\end{eqnarray*}
and let $(T_d)$ be a sequence of $d \times d$ nonnegative definite Hermitian matrices. Write $\Z_n = (\mathbf{u}_1, \cdots, \mathbf{u}_n)$ where $\mathbf{u}_j = (w_{1j}, \cdots, w_{dj})^{T}$ and $\X_n = T_d^{\frac{1}{2}}\Z_n$. Then, the sample covariance matrix of $\X_n$ can be expressed as $\Sbf_n = \frac{1}{n}\X_n \X_n^{T} = \frac{1}{n} T_d^{\frac{1}{2}}\Z_n \Z_n^{T}T_d^{\frac{1}{2}}$. Let $(H_n)$ be a sequence of the empirical spectral distributions of $(T_d)$ with the dimension to sample size ratio, $y_n = \frac{d}{n}$, and assume that $H_n$ weakly converges to a nonrandom probability distribution $H$ on $[0, \infty)$ as $n$ tends to infinity satisfying $y_n \rightarrow y$ where $y$ is a positive constant. The limit distribution $H$ is referred to as the population spectral distribution (PSD). For the simplest case where $T_d = I_d$, the ESDs of $(T_d)$ are $H_{n}(t) = \delta_{1}(t)$ for all $n$ and thus the PSD $H(t)$ can be obtained as $H_n (t) \rightarrow H(t) = \delta_{1}(t)$. The classical Marcenko-Pastur law says that the ESD of $\Sbf_n$ converges to a nonrandom limiting spectral distribution (LSD) $G$ under this case.

A generalized version of the classical Marcenko-Pastur law has been developed \citep*{silverstein1995strong} when the PSD $H(t)$ is an arbitrary probability measure. The PSD $H$ and the LSD $G$ are linked by the following inverse map of the companion Stieltjes transform $\underline{s}(z)$ of the LSD $G$,
\begin{eqnarray*} 
z = g_{y,H}(\underline{s}) = - \frac{1}{\underline{s}} + y \int{\frac{t}{1+t\underline{s}}}dH(t),~~~~z \in \mathbb{C}^{+},
\end{eqnarray*}
which is called the Silverstein equation. Let $F_{y,H}$ be the distribution whose Stieltjes transform is $m_{y,H}=g_{y,H}^{-1}$. Throughout the proposal, we call $F_{y,H}$ the Marcenko-Pastur (MP) distribution with indexes $(y, H)$. A lot of theory and applications in the spectral analysis have been established from this equation. One crucial result is the Lemma \ref{lem:silverstein} which indicates the analytical relationship between the two supports of the PSD $H$ and the MP distribution $\MP$ \citep*{silverstein1995analysis}. Define 
\begin{eqnarray} \label{eq:silverstein}
\psi_{y,H} (\alpha) = g_{y,H}(-1/\alpha) = \alpha + y \alpha \int{\frac{t}{\alpha-t} dH(t)} 
\end{eqnarray}
for $\alpha \notin \Gamma_{H}$ and $\alpha \neq 0$. 
\begin{lemma}\label{lem:silverstein} (Silverstein and Choi, 1995) 
If $\lam \notin \MPsupp$, then $\MPm (\lam) \neq 0$ and $\alpha = -1/\MPm(\lam)$ satisfies
\begin{enumerate}
\item[(a)] $\alpha \notin \Hsupp$ and $\alpha \neq 0$ (so that $\psi_{y,H}(\alpha)$ is well-defined);
\item[(b)] $\psi_{y,H}^{'}(\alpha) > 0$.
\end{enumerate}
Conversely, if $\alpha$ satisfies (a)-(b), then $\lam = \psi_{y,H}(\alpha) \notin \MPsupp$.
\end{lemma}
\begin{figure}[t]
	\centering
	\includegraphics[scale=0.5]{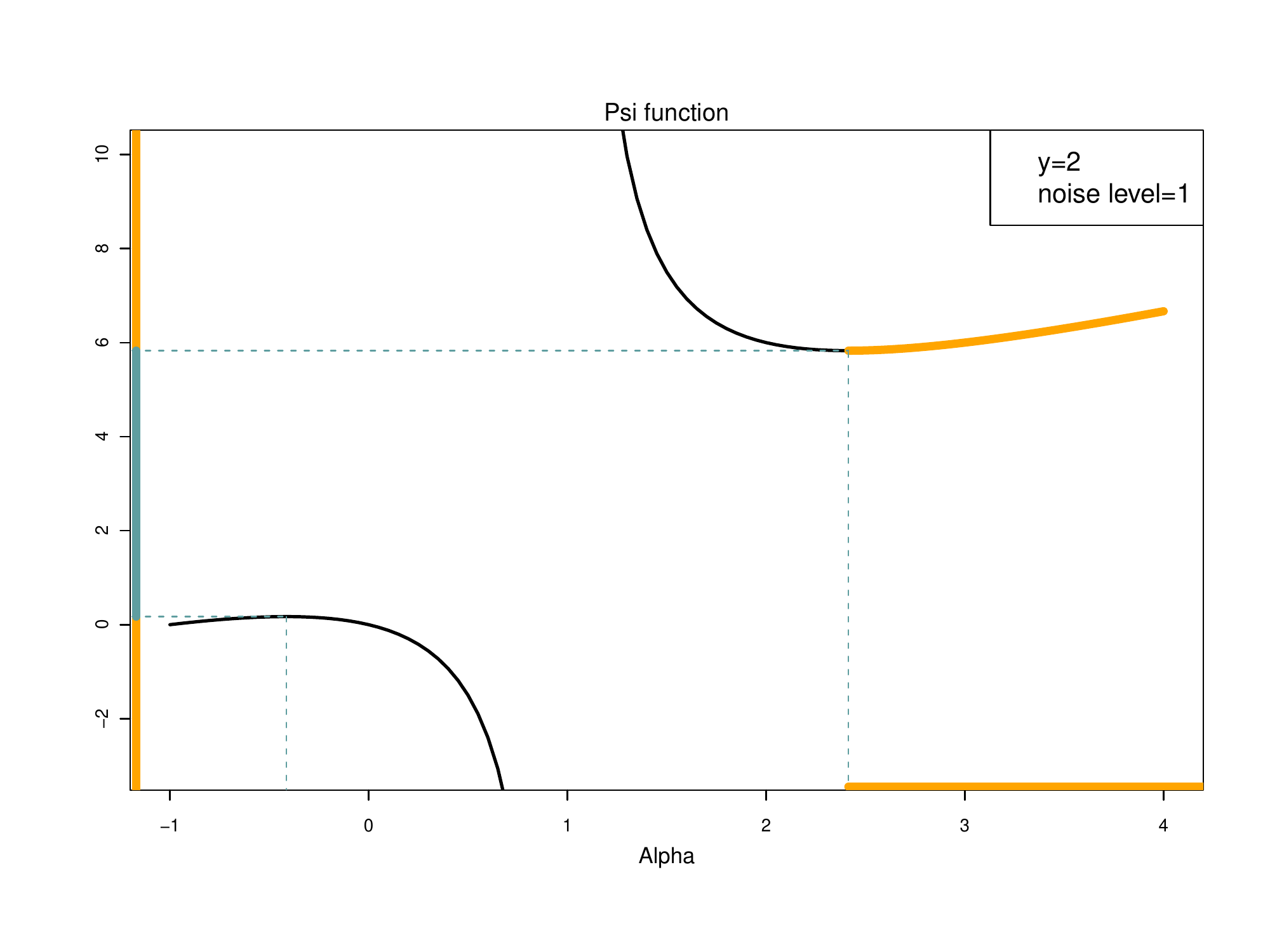}
	\caption{The curve illustrates $\MPpsi$ with $y=2$ and the PSD $H=\delta_{1}$. The yellow regions correspond to the $\alpha$ satisfying (a) and (b) in Lemma \ref{lem:silverstein} and the blue region indicates the support of the MP distribution $\MP$.}
\label{fig:psi_function}
\end{figure}
Roughly speaking, what the lemma says is that given the PSD $H$ one can characterize the support of the MP distribution $\MP$ as illustrated in Figure \ref{fig:psi_function}. In Figure \ref{fig:psi_function}, the curve indicates the $\MPpsi(\alpha)$ with $y=2$ and the PSD $H=\delta_{1}$ and the regions indicated by yellow are $\{ \alpha \}$ satisfying (a) and (b) in Lemma \ref{lem:silverstein} and the corresponding $\MPpsi(\alpha)$. According to Lemma \ref{lem:silverstein}, the support of the MP distribution $\MP$ is indicated as blue on the y-axis.     

\cite{baik2006eigenvalues} investigated the asymptotic behaviors of the sample eigenvalues corresponding to the spikes under Johnstone's spike model. They showed that sample spike eigenvalues converge almost surely to some functions of the corresponding population spike eigenvalues  and the different functions are considered for the different types of spikes. \cite{bai2012sample} extended their results to a generalized spike model where the spectrum of a base set follows an arbitrary distribution. A generalized spike model has a population covariance matrix $T_d$ that is of the form
\begin{eqnarray*}
T_d = 
\left( \begin{array}{cc}
\Lam & 0 \\
0          & V_d  
\end{array} \right)
\end{eqnarray*}
where $\Lam$ is an $M \times M$ matrix whose eigenvalues are the population spikes, $\alpha_1> \cdots> \alpha_K$ of respective multiplicity ($n_k$) with $\sum_{k=1}^{K}n_k=M$, and $V_d$ is a $d' \times d'$ matrix whose eigenvalues are $\beta_{n,1} \geq \cdots \geq \beta_{n,d'}$. Under certain conditions on the true eigenvalues, they showed the almost sure convergence of a sample spike eigenvalue to some function of the corresponding population eigenvalue as $n$ and $d$ both increase such that $\frac{d}{n} \rightarrow y$. Our method is motivated by a generalized spike model, which allows flexibility on non-spike population eigenvalues.

\section{Methodology}\label{section:method}
In this section, we propose an algorithm to choose the number of spikes for a generalized spike model where the PSD is not constrained to be the point mass at $\sigma^2$, i.e. $H(t)=\delta_{\sigma^2}(t)$. The algorithm will be first described under the scenario where the PSD is known, which is unrealistic in most cases but easy to understand the basic idea. Next, we will introduce the main algorithm which is applicable when the PSD is unknown. Also, we propose a graphical tool to assess the distribution assumption for the PSD. In this section, we denote the PSD by $\PSD$ to emphasize the parameters that identify the $H(t)$ and the eigenvalues of $\Sn$ in decreasing order by $\hat{\lambda}_1, \cdots, \hat{\lambda}_{d \wedge n}$.

\subsection{Estimation when the PSD is known}

Let us first consider the case when the PSD $\PSD$ is known. Since we know the support of the corresponding LSD $\MP$ according to Lemma \ref{lem:silverstein}, a straightforward way to estimate the number of underlying spikes may be counting the number of eigenvalues above the upper boundary of the support. However, this procedure may slightly overestimate the true number of spikes because it ignores the variation in the largest noise eigenvalue. In the point mass PSD, for example, the resulting upper boundary for the LSD is $b_y=\sigma^2 (1+\sqrt{y})^2$ whereas the largest eigenvalue is known to converge in distribution to the Tracy-Widom law whose support contains $b_y$, giving a non-ignorable probability of the largest eigenvalue being greater than $b_y$ \citep{johnstone2001distribution,ma2012}. To reflect such variation of the largest noise eigenvalue, in this particular example, the approximate Tracy-Widom quantile can replace the upper boundary $b_y$ and this provides a more precise estimation \citep{kritchman2008determining}.

However, except for this simplest case ($\PSD = \delta_{\sigma^2}$), we do not know the distribution of the largest noise eigenvalue. Thus, we approximate the distribution by simulation under a sufficiently large number of independent replications and then we take a certain quantile (e.g. 99th percentile) as a threshold above which sample eigenvalues are considered as spikes. The simulation procedure to get a level $\alpha$ threshold, $s_{\alpha}$, is described in Algorithm 1.

\textbf{Algorithm 1.}
\begin{enumerate}
    \item For $b=1, 2, \cdots, B$ with a sufficiently large number $B \in \mathbb{N}$, 
\begin{enumerate}
    \item Generate $d$ random variables $\{ \beta_{1}^{(b)}, \cdots, \beta_{d}^{(b)} \}$ from $\PSD$ and get a $d \times d$ diagonal matrix $T_{(b)}$ by taking $\mathrm{diag}(T_{(b)}) = \{ \beta_{1}^{(b)}, \cdots, \beta_{d}^{(b)} \}$;
    \item Generate a $d \times n$ matrix $Z_{(b)}$ whose entries are independent variables from $N(0, 1)$;
    \item Get the largest eigenvalue $\hat{\lambda}_{1}^{(b)}$ of $\frac{1}{n}T_{(b)}^{1/2}Z_{(b)}Z_{(b)}^{T}T_{(b)}^{1/2}$.
\end{enumerate}
    \item Obtain $(1-\alpha)$-quantile $s_{\alpha}$ based on the set $\{ \hat{\lambda}_{1}^{(1)}, \cdots, \hat{\lambda}_{1}^{(B)} \}$.
\end{enumerate}

\subsection{Estimation when the PSD is unknown}

In practice, the PSD $\PSD$  is unknown and should be estimated as well. Here we consider the scenario that the type of distribution is known but with unknown parameters, e.g. the case that the PSD is $\delta_{\sigma^2}$ with an unknown parameter $\sigma^2$. The main algorithm is based on a sequence of nested hypothesis tests: 
\begin{eqnarray*}
    H_0^{(m)}  : K \leq m-1 ~~~~~ vs. ~~~~~ H_1^{(m)}  : K \geq m
\end{eqnarray*}
where $K$ is the true number of spikes and $m=1, \cdots, d \wedge n $. At the $m$-th stage, we estimate the PSD parameters assuming $\hatlam_m, \cdots, \hatlam_{d \wedge n}$ to be non-spikes and test whether or not the $\hatlam_m$ is from a spike based on the approximate distribution of the largest noise eigenvalue obtained by Algorithm 1 with the estimated parameters. If the null is rejected, i.e. there are at least $m$ spikes, then we proceed to the $(m+1)$-th hypothesis test after excluding $\hatlam_m$. Otherwise, we stop the procedure and conclude that there are at most $m-1$ spikes. Note that when we consider a point mass PSD and a theoretically obtained threshold from the Tracy-Widom law instead of the simulated one, this procedure plays the same role as the method proposed by Kritchman and Nadler \citep{kritchman2008determining} with a carefully estimated noise variance. 

\textbf{\textit{Estimation of unknown parameters of the PSD.}}\indent An interesting topic in the random matrix theory is the estimation of the PSD parameters. \citep*{li2013estimation,bai2010estimation}. Although we employ the Bai's method of moments in this paper because it is intuitive and is to implement, other PSD parameter estimation methods can be used as well. \cite{bai2010estimation} proposed a method to estimate the unknown parameters based on the relationship between the moments of the PSD $\PSD$ and the moments of the MP distribution $\MP$. The following lemma \citep*{nica2006lectures} describes the relationship. 
\begin{lemma}\label{lem:moments} (Nica \& Speicher, 2006) The moments $\alpha_j = \int x^j dF_{y,H}(x),~j \geq 1$ of the LSD $F_{y,H}$ are linked to the moments $\beta_j = \int t^j dH(t)$ of the PSD $H$ by 
\begin{eqnarray}\label{eq:sample_population_moments}
\alpha_j = y^{-1}\sum y^{i_1 + i_2 + \cdots + i_j} (\beta_1)^{i_1}(\beta_2)^{i_2}\cdots (\beta_j)^{i_j} \phi^{(j)}_{i_1,i_2,\cdots,i_j}
\end{eqnarray}
where the sum runs over the following partitions of $j$:
\begin{eqnarray*}
(i_1, \cdots, i_j) : j = i_1 + 2i_2 + \cdots + ji_j , ~~~~i_l \in \mathbb{N},
\end{eqnarray*}
and $\phi^{(j)}_{i_1,i_2,\cdots,i_j}$ is the multinomial coefficient
\begin{eqnarray*}
\phi^{(j)}_{i_1,i_2,\cdots,i_j} = \frac{j!}{i_1! i_2! \cdots i_j! (j+1-(i_1 + i_2 + \cdots + i_j))!}.
\end{eqnarray*}
\end{lemma}
We denote the estimate of $\bm{\theta}$ obtained from the method of moments by $\hat{\bm{\theta}}$. Note that the proposed moment estimator has been proved to be strongly consistent and asymptotically normal \citep{bai2010estimation}. 

We employ this method of moments to estimate the PSD parameters $\bm{\theta}$ and the main algorithm to estimate the number of spikes $K$ is described as follows: \\
\textbf{Main algorithm.} \\
From $m=1$, iterate the following procedure until it stops.
\begin{enumerate}
    \item Based on $\{ \hatlam_{m}, \hatlam_{m+1}, \cdots, \hatlam_{d \wedge n} \}$, obtain the PSD parameters $\hat{\bm{\theta}}^{(m)}$. 
    \item Applying Algorithm 1 with the estimated $\hat{\bm{\theta}}^{(m)}$, obtain the $(1-\alpha)$-quantile $s_{\alpha}^{(m)}$ with a predetermined level $\alpha$. 
    \item If $\hatlam_m > s_{\alpha}^{(m)}$, reject $H_{0}^{(m)}$ and go to the step 1 replacing $m$ by $m+1$. Otherwise, we do not reject $H_0^{(m)}$, the estimate of the number of spikes is $\hat{K}=m-1$, and stop here. 
\end{enumerate}

\subsection{PSD diagnostics}\label{subsection:PSD_diagnostics}

The proposed method for determining the number of spikes depends on the correct specification of the underlying PSD. Under a misspecified PSD, the estimated number would be unreliable. Here, we develop a graphical tool for the diagnostic to check if the assumed PSD is correctly specified. 

Let $\underbar{s}$ be the true companion Stieltjes transform as defined in Section \ref{section:RMT} and let the sample companion Stieltjes transform, denoted by $\underbar{s}_n$,  be 
\begin{eqnarray*}
    \underbar{s}_n (u)= -\frac{1-d/n}{u} + \frac{1}{n}\sum_{i=1}^{d} \frac{1}{\hatlam_i - u}
\end{eqnarray*}
for $u \in \Gamma_{F_n}^{c}$ where $F_n$ denotes the ESD from the data and $\Gamma_{F}$ denotes the support of the distribution $F$. Then, (reference - weiming) shows that, under certain conditions, (a) $\underbar{s}_n (u)$ converges to $\underbar{s}(u)$ for any $u \in  \mathrm{int}(\lim \inf_{n \rightarrow \infty} \Gamma_{F_n}^{c} \backslash 0)$; (b) the $\psi_{y,H}(\alpha)$ defined in (\ref{eq:silverstein}) uniquely determines the PSD $H$. Combining (a) and (b) gives us an important fact that if the PSD of the data is truely $H$, then the $\hat{\psi}_n(\alpha) = \underbar{s}_n^{-1}(-1/\alpha)$ will be close to the true $\psi_{y,H}(\alpha)$ for $\alpha \in A$ where $A = \{\alpha: \alpha \notin \Hsupp, \alpha \neq 0, ~\mathrm{and}~\psi_{y,H}^{'}(\alpha) > 0 \}$.  

Based on this fact, we propose \textit{psi envelopes} which carefully study the difference between $\psi_{y,H}$ and $\hat{\psi}_n$ by constructing simulated envelopes, $\hat{\psi}_n^{q}$ for $q=1, \cdots, Q$. The basic idea is the same for Q-Q envelopes \citep{hannig2001zooming}. The envelope $\hat{\psi}_n^{q}$ is from $Q$ independent replications obtained as in Algorithm 1. Then, the psi envelope enables assessment of the PSD assumption by checking whether or not the $\hat{\psi}_n$ is covered by the envelope. 
\begin{figure}[t]
	\centering
	\includegraphics[width=15cm]{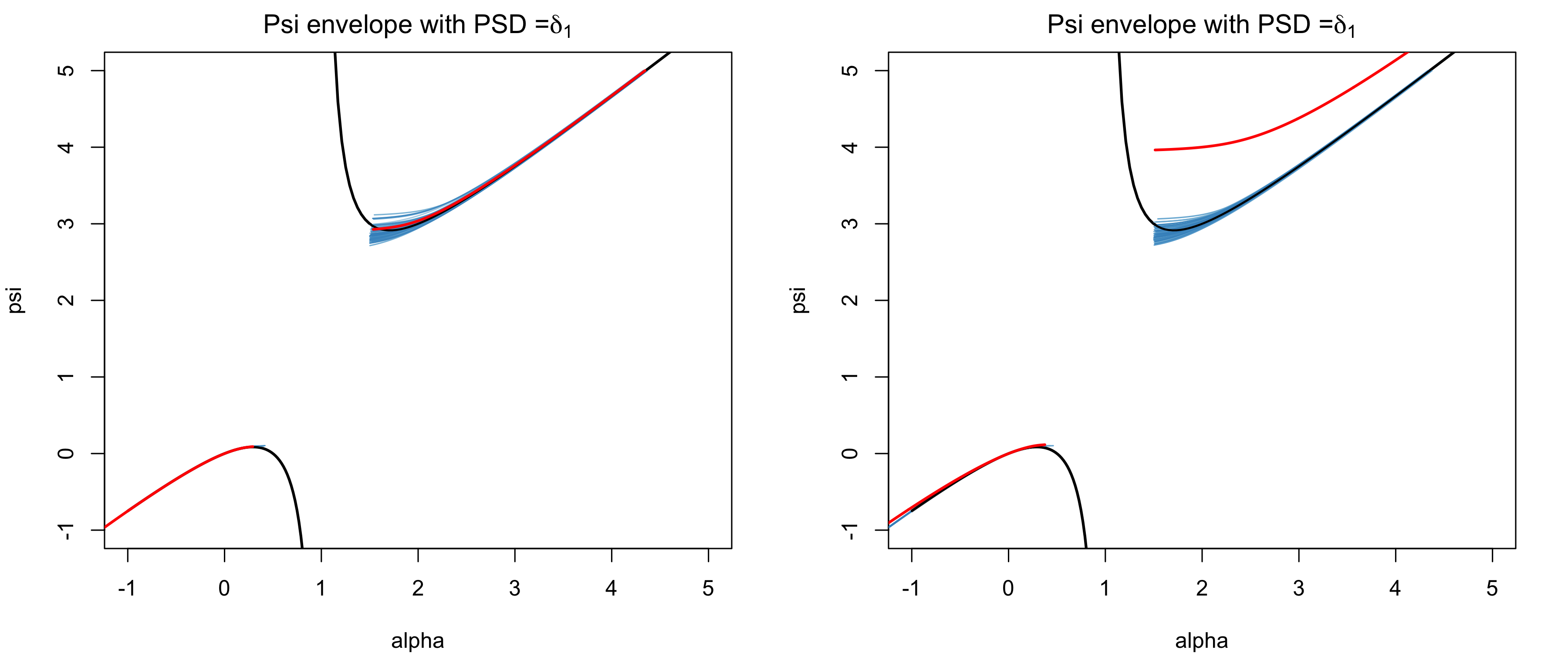}
	\caption{Examples of the psi envelope for assessment of the point mass PSD $H=\delta_{1}$. The left panel shows the case where the sample eigenvalues are truly from the assumed $H$ whereas the right panel shows the case where the sample eigenvalues are from the different point mass PSD $H^{'}=\delta_{1.2}$. The figure demonstrates the sensitivity of the psi envelope diagnostic.}
\label{fig:Psi_envelope}
\end{figure}
Two examples of the psi envelope are shown in Figure \ref{fig:Psi_envelope}, which are checking whether or not two different sets of eigenvalues are from the PSD $H=\delta_{1}$. For the left plot, the sample eigenvalues truly from the $H=\delta_{1}$ are considered, and the sample eigenvalues from a different PSD $H^{'}=\delta_{1.2}$ for the right plot. In each plot, the function $\psi_{y,\delta_{1}}$ is shown in black with $100$ blue envelope and the estimated function $\hat{\psi}_n$ in red. The $\hat{\psi}_n$ which is based on the eigenvalues truly from the $H$ is well covered by the envelope whereas the $\hat{\psi}_n$ from $H^{'}$ shows clear deviation from the envelope. As this example shows, the psi envelope provides a useful graphical tool for the PSD diagnostic.

\section{Application to RNA-seq data}\label{section:application}

\begin{figure}[t]
	\centering
	\includegraphics[width=12cm]{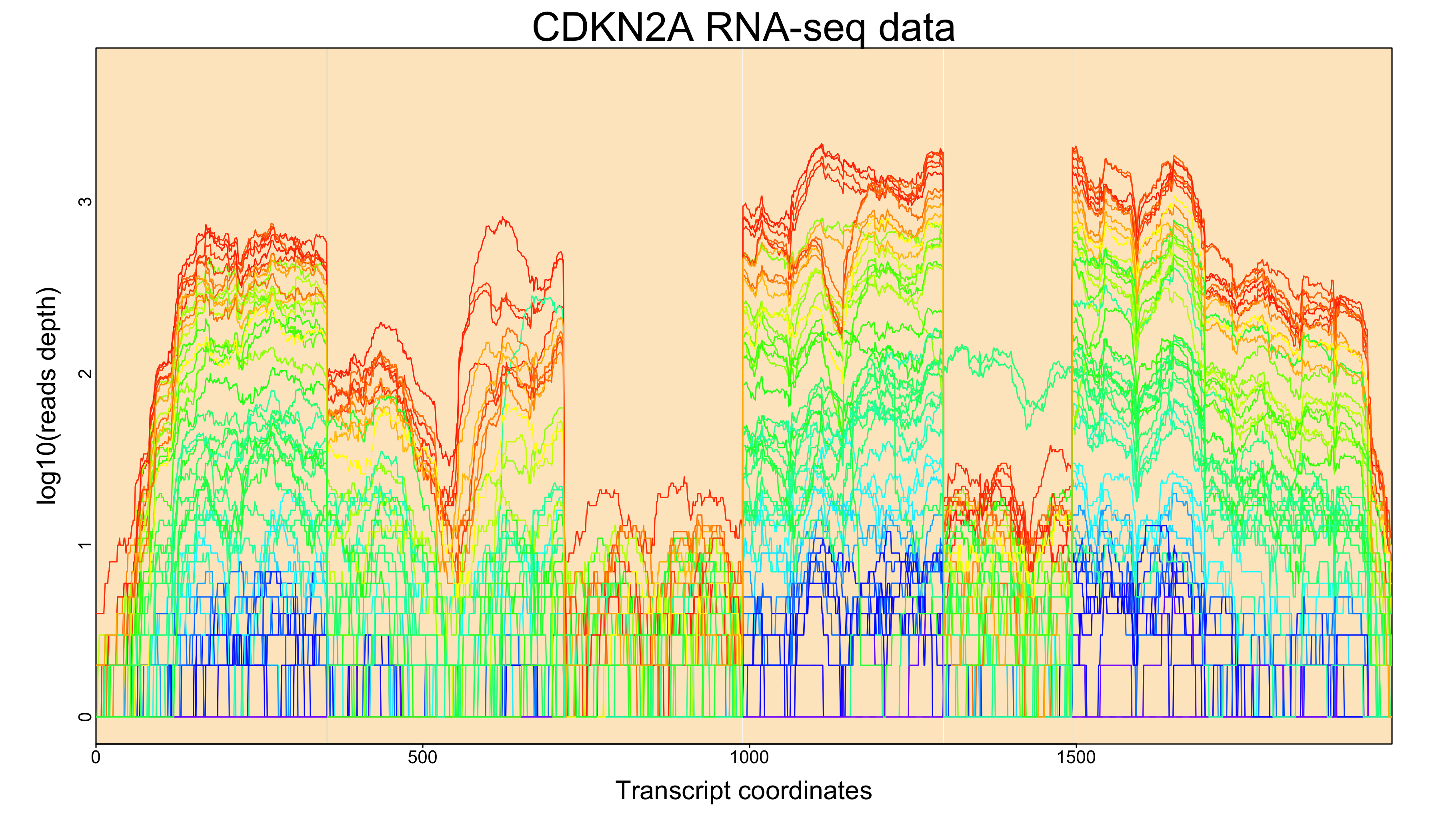}
	\caption{A set of RNA-seq observations for the gene CDKN2A are plotted on the log scale.}
\label{fig:CDKN2A_overlay}
\end{figure}

In this section, we apply the proposed methods to our main example, RNA-seq gene expression data. Let us first briefly describe the data structure. A collection of 522 head and neck squamous carcinoma RNA-seq observations were obtained from the TCGA Research Network and the dataset wes processed as described in The Cancer Genome Atlas Research Network (2012). In this paper, we study the base-position gene expression for each of several important cancer related genes.
For each gene, the RNA-seq read-depths are measured along the length of the transcript. The resulting data structure is an expression count matrix $\bm{R}=\{ r_{ij} \}_{1\leq i \leq d, 1 \leq j \leq n}$ where $r_{ij}$ is the read count at the $i$th position for the $j$th patient, $d$ is the length of the transcript at the gene being studied, and $n$ is the sample size. Since RNA-seq counts data show unstable variations within and between observations, the shifted logarithm transformation was taken to stabilize such heterogeneity, i.e. $\X = \{ x_{ij} \}$ where $x_{ij}=\log_{10}(r_{ij}+1)$. For each gene, our analysis is based on the resulting matrix $X$ and the example of the gene CDKN2A is described in Figure \ref{fig:CDKN2A_overlay}.

Since RNA-seq data typically have 1,000-20,000 dimensions while there are only hundreds of samples available, PCA is a very useful tool to reduce a huge dimension size and visualize the underlying relationship between samples or variables. In many cases, RNA-seq data are analyzed for hundreds or thousands of genes for various purposes such as discovery of key genes, detection of interesting genetic events, or identification of novel clusters, so that an automatic choice of the number of PCs at each gene is useful. To our best knowledge, however, there is no existing method to select the number of PCs that is applicable to RNA-seq data. In Section \ref{subsection:proposed_noise_model}, we show why the existing methods are not appropriate for the RNA-seq data and suggest a new PSD model which is more suitable. And we compare our results with some existing methods for several important genes in Section \ref{subsection:important_genes}. 

\subsection{The proposed noise model}\label{subsection:proposed_noise_model}

\begin{figure}[t]
	\centering
	\includegraphics[width=15cm]{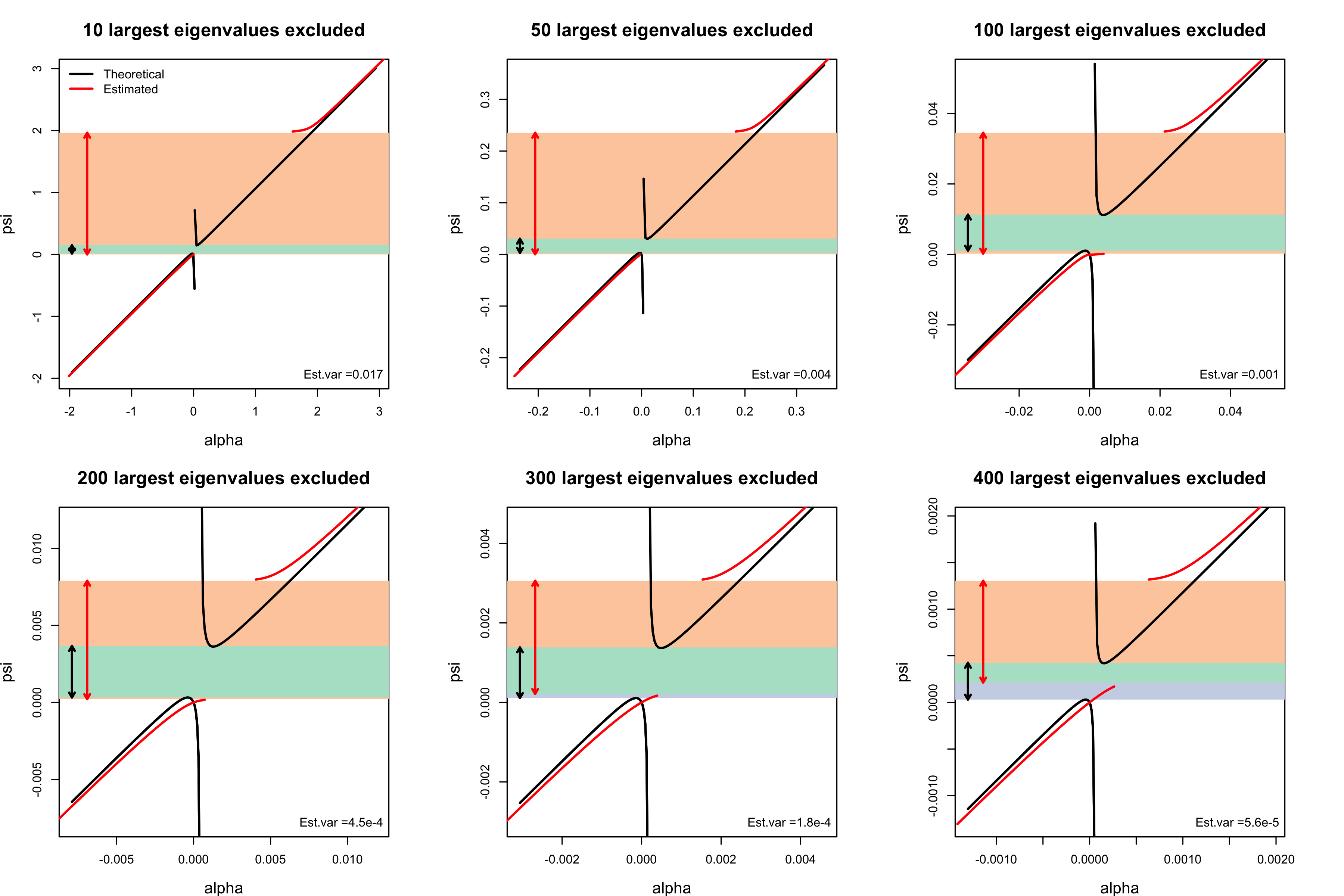}
	\caption{Comparison between the theoretical functions $\psi_{y_n, \delta_{\hat{\sigma}^2}}$ (black) and the estimated functions $\hat{\psi}_n$ (red) for the RNA-seq data at the gene CDKN2A. In each plot, the estimates $\hat{\sigma}^2$ and $\hat{\psi}_n$ are obtained based on the sample eigenvalues except for the 10, 50, 100, 200, 300, 400 largest eigenvalues. The blue rectangular area with the black arrows indicate the theoretically expected supports of the LSDs under the point mass PSD. The orange rectangular area with the red arrows indicate the supports of the ESDs from the data. The green rectangular area indicates the intersection of the blue and orange regions. The estimated noise variances are provided at the bottom of each plot. The Figure shows a point mass PSD provides a very poor fit to the data.}
\label{fig:CDKN2A_white_noise_envelopes}
\end{figure}

Figure \ref{fig:CDKN2A_white_noise_envelopes} shows graphics which demonstrate that the point mass PSD for a noise distribution of RNA-seq data is clearly not appropriate. Here, the gene CDKN2A is considered with $d=1978,~n=522$. In each plot, the theoretical function $\psi_{y_n, \delta_{\hat{\sigma}^2}}$ (black) with the estimated noise level $\hat{\sigma}^2$ and $y_n=\frac{d}{n}$ is compared with the estimated function $\hat{\psi}_n$ from data (red) after excluding the 10, 50, 100, 200, 300, 400 largest eigenvalues sequentially. The red arrow represents the support of the empirical spectral distribution from the data, that is, from the smallest to the largest sample eigenvalue. Correspondingly, the support of the theoretical M-P distribution extended to the 99th percentile of the distribution of the largest eigenvalue, known as the Tracy-Widom law, is represented by the blue arrow \citep{ma2012}. This enables more accurate comparisons of the two distributions by taking into account the variation in the maximum eigenvalue. Although the theoretical and data-driven functions, $\psi_{y_n, \delta_{\hat{\sigma}^2}}$ and $\hat{\psi}_n$, as well as the corresponding supports become comparable as more eigenvalues are kicked out, it is clear that they strongly disagree even when almost all eigenvalues are elliminated. This demonstrates that the noise eigenvalues from the data do not follow the classical M-P distribution, which motivates our improved PSD models.

Under the questionable white-noise assumption, we easily expect that too many PCs would be determined to be significant because of the extreme skewness of the ESD. The severe positive skewness makes the $k$th sample eigenvalue, the maximum of the remaining eigenvalues at the $k$th stage, be much greater than the theoretically possible maximum eigenvalue, resulting in the rejection of the hypothesis that the $k$th eigenvalue is from noise. In almost all cases, we observed that this was indeed true as described in Table \ref{tab:genes_results}. From the perspective of the dimension reduction, however, such high numbers of PCs may not be helpful especially for downstream statistical analyses that mostly require a much smaller dimension size. 

\begin{figure}[t]
	\centering
	\includegraphics[width=15cm]{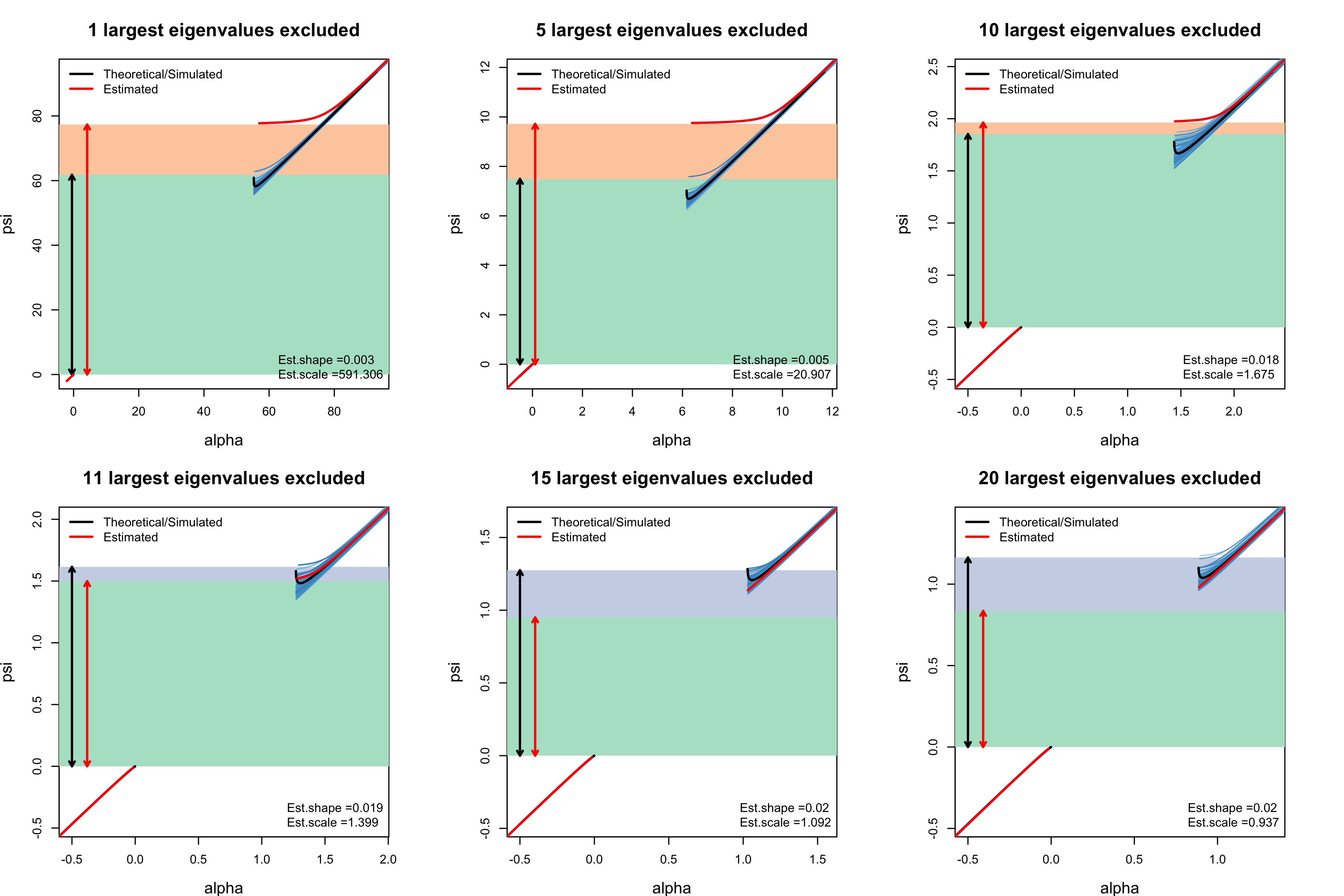}
	\caption{The psi envelopes for assessing the Gamma PSD for the gene CDKN2A. In each plot, the remaining eigenvalues excluding the 1, 5, 10, 11, 15, 20 largest eigenvalues are compared with the estimated Gamma PSD. The blue, orange, and green rectangular regions and red and black arrows are determined similarly as in Figure \ref{fig:CDKN2A_white_noise_envelopes}. The PSD parameters are provided at the bottom of each plot. The figure indicates 11 large non-noise eigenvalues.}
\label{fig:CDKN2A_gamma_noise_envelopes}
\end{figure}

Accordingly, the PSD that has a point mass at the noise variance, i.e. $H(t)=\delta_{\sigma^2}$, is not a suitable distribution for RNA-seq data whose eigenvalues are extremely right-skewed. To capture the extreme positive skewness, we suggest a right-skewed PSD, particularly the Gamma distribution truncated at some upper quantile. Other right-skewed distributions may be used but the Gamma distribution fit the best from our experience. Also, we can control the degree of positive skewness by adjusting the truncation quantile. Our method is sensitive to this choice of the quantile and precise specification is a  topic for future research. Truncation determined by the upper 0.995 quantile gave reasonable values so we use it for the rest of the paper. Figure \ref{fig:CDKN2A_gamma_noise_envelopes} shows the psi envelopes introduced in Section \ref{subsection:PSD_diagnostics} for assessing the gamma assumption after kicking out a few largest sample eigenvalues. As in Figure \ref{fig:CDKN2A_white_noise_envelopes}, Figure \ref{fig:CDKN2A_gamma_noise_envelopes} shows the estimated functions $\hat{\psi}_n$ in red sequentially removing a first few eigenvalues with the red vertical arrows for the supports of the ESDs. The black curves represent the theoretical functions $\psi_{y_n, H(t; \hat{\bm{\theta}})}$ with the black arrows indicating the approximate supports of the corresponding LSDs. As more eigenvalues are excluded up to 11, the sample psi curve gets closer to the psi envelopes, which supports that the remaining sample eigenvalues roughly follow the LSD based on the estimated PSD $H(t,\hat{\bm{\theta}} )$. As we will see in Section \ref{subsection:important_genes}, the critical value 11 is exactly the estimated number of spikes from the proposed method in Section \ref{section:method}. 

\subsection{Application to important genes} \label{subsection:important_genes}

We compared our method (CM) with the methods proposed by Kritchman and Nadler (KN) and Passemier and Yao (PY) for the RNA-seq data at several important tumor-related genes: CDKN2A, TP53, FAT1, PTEN, CASP4, CHEK2, EGFR, and PIK3CA. Let us first briefly describe the two methods.  

\textbf{Method of Kritchman and Nadler (KN).}
\cite{kritchman2008determining} developed an algorithm for rank determination based on the asymptotic distribution of the largest noise eigenvalue. Their algorithm performs sequential hypothesis tests on whether the largest eigenvalue at each step arises from a signal rather than from noise. The statistical procedure at each step involves estimating noise variance and setting a threshold based on the Tracy-Widom distribution where the largest noise eigenvalue follows \citep*{johnstone2001distribution}. Their algorithm has been considered as a good benchmark for judging performance of other methods for determining the number of components in many papers. 

\textbf{Method of Passemier and Yao (PY).}
\cite{passemier2012determining,passemier2014estimation} proposed a method for estimating the number of spikes under the case where there are possibly equal spikes. Based on the different asymptotic behaviors of spike and non-spike sample eigenvalues, they determine a threshold for the successive spacings of the ordered sample eigenvalues. Because larger spacings are expected for spike sample eigenvalues than for noise eigenvalues, one may separate spikes and non-spike eigenvalues based on an appropriately determined threshold for the spacing. This method is very intuitive in the sense that the proposed procedure is somehow similar to the naive procedure based on scree plots with a more reasonable separation based on random matrix theory. To avoid false determination due to ties of spike eigenvalues, they also proposed a more robust estimator by using consecutive two or more spacings that should be larger than a threshold at the same time to be considered as spikes. 

\begin{table}[tb]
\centering
\begin{tabular}{r R{1.5cm} R{2.7cm} R{2.7cm} R{2.7cm}} \toprule
    & d & CM $(\hat{\tau},~\hat{\nu})$ & KN ($\hat{\sigma}^2$) & PY ($\hat{\sigma}^2$) \\ \midrule
    CDKN2A & 1978 & 11 (0.019, 1.399) & 521 ($\mathrm{NA}$) & 197 (0.00046) \\
    CHEK2 & 2595& 5 (0.014, 2.029) & 521 ($\mathrm{NA}$) & 208 (0.00034) \\
    CASP4 & 2688 & 6 (0.011, 3.665) & 521 ($\mathrm{NA}$) & 219 (0.0039) \\
    PIK3CA & 3686 & 8 (0.013, 1.310) & 521 ($\mathrm{NA}$) & 521 ($\mathrm{NA}$) \\
    TP53 & 3876 & 12 (0.013, 1.505) & 521 ($\mathrm{NA}$) & 521 ($\mathrm{NA}$) \\
    PTEN & 5535 & 17 (0.011, 1.235) & 521 ($\mathrm{NA}$) & 521 ($\mathrm{NA}$) \\
    EGFR & 7965 & 13 (0.010, 2.434) & 521 ($\mathrm{NA}$) & 521 ($\mathrm{NA}$) \\
    FAT1 & 15232 & 16 (0.008, 3.781) & 521 ($\mathrm{NA}$) & 521 ($\mathrm{NA}$)    \\ \bottomrule
\end{tabular}
\caption{Estimates of the number of spikes from the proposed method (CM) and two existing methods (KN and PY). The lengths of transcripts (dimensions of the RNA-seq data) for eight genes are provided in the second column. For the CM method, the estimated shape and rate parameters $(\hat{\tau},~\hat{\nu})$ of the Gamma are provided and the estimated noise variance $\hat{\sigma}^2$ are also provided for the KN and PY methods. The table shows that the proposed CM method estimates reasonable numbers of spikes whereas the KN and PY methods provides far too large number of spikes. NA indicates that the noise variance is not available.}
\label{tab:genes_results}
\end{table}
The estimated number of spikes for each gene from the three methods are summarized in Table \ref{tab:genes_results}. As expected, both KN and PY methods determine a huge number of spikes and, in particular, the KN results indicate that all PCs are spikes for these eight genes. When the all PCs are declared as spikes, the noise variance cannot be estimated because there is no noise eigenvalue any more, as indicated by NA in Table \ref{tab:genes_results}. On the other hand, our proposed method provides biologically reasonable and practical number of spikes. Although we provide the results for the eight chosen genes, we have observed that this is indeed true for almost all genes. We believe that the proposed method can give valuable contribution to distinguish meaningful and important signals from noise in RNA-seq data. Furthermore, the method can be also applied to other types of data set where a point mass PSD is not appropriate with a carefully chosen PSD. 

\pagebreak
\bibliographystyle{apalike}
\bibliography{references} 
\end{document}